\documentclass[twocolumn]{jetpl}
\usepackage[english]{babel}
\usepackage[utf8]{inputenc}
\usepackage[T1]{fontenc}
\usepackage{cite}
\usepackage{mathtools}
\usepackage{xcolor}

\usepackage{amsmath}
\allowdisplaybreaks
\usepackage{cuted}
\usepackage{subcaption} 
\usepackage{epstopdf}
\usepackage{float}
\usepackage{multirow}
\usepackage{caption}

\title{Magnon-polaritons in terahertz resonators based on antiferromagnets with weak ferromagnetism}

\rtitle{Magnon-polaritons in THz cavities}

\sodtitle{Magnon-polaritons in terahertz resonators based on antiferromagnets with weak ferromagnetism}

\author{
A.A. Luzgina$^{1,2}$, A.A. Kraev$^{1,2}$, S. Cao$^{3}$, A.K. Zvezdin$^{1}$, V.I. Belotelov$^{1,4}$
}

\rauthor{Luzgina, Kraev, Cao, Zvezdin, Belotelov}

\sodauthor{A.A. Luzgina$^{1,2}$, A.A. Kraev$^{1,2}$, S. Cao$^{3}$, A.K. Zvezdin$^{1}$, V.I. Belotelov$^{1,4}$}

\address{~\\$^1$Russian Quantum Center, Skolkovo Innovation City, 121205 Moscow, Russia\\~\\$^2$ Moscow Institute of Physics and Technology (National Research University), 141701 Dolgoprudny, Russia\\~\\$^3$ Department of Physics, Shanghai University, Shanghai, China\\~\\$^{4}$ Lomonosov Moscow State University, 119991 Moscow, Russia}

\dates{}{*}

\abstract{Terahertz magnon-polaritons in antiferromagnets with weak ferromagnetism are theoretically investigated using YFeO$_3$ as an example. Dispersion relations are obtained for the quasi-ferromagnetic and quasi-antiferromagnetic modes. It is shown that the intrinsic Fabry--Perot resonances of the magnetic slab provide strong coupling, but their low quality factor limits the cooperativity. A photonic crystal cavity (PhCC) increases the cooperativity of the quasi-antiferromagnetic mode from $C\sim6$ to $C\sim41$ at 300~K and to $2.2\times10^3$ at 63~K.}

\renewcommand{\thetable}{\Roman{table}}
\begin{document}
	
	\maketitle
	\textbf{1. Introduction.}\\
Magnon-polaritons are hybrid excitations arising from the resonant interaction of collective spin excitations with electromagnetic waves. The theory of coupled magnon-photon modes in antiferromagnets was developed as early as the early 1970s \cite{manohar1972,bose1975}, and the first experimental evidence of magnetic polaritons was obtained in studies of antiferromagnetic resonance in $\mathrm{FeF_2}$ in the far-infrared range \cite{sanders1978}.

In recent years, interest in magnon-polaritons has increased significantly owing to the development of studies of strong coupling between magnons and electromagnetic modes of cavities \cite{rameshti2022}. The combination of photonic and magnonic components opens up possibilities for coherent energy conversion between the electromagnetic and spin subsystems, control of electromagnetic signals, and their delay and storage \cite{wolz2020,wang2021}. Magnon-photon systems are also of interest for quantum technologies, including the transfer and storage of quantum information and the integration of magnons with other quantum systems \cite{shen2025}. For such tasks, the high-cooperativity regime is particularly important, in which the coherent interaction between the magnon and photon modes prevails over dissipative losses.

Antiferromagnets are especially attractive for magnon-polaritonics owing to the high eigenfrequencies of their spin excitations, which reach the terahertz range. An important class of such systems is formed by rare-earth orthoferrites $\mathrm{RFeO_3}$ \cite{belov1979,turov2001,kurihara2023}, which are antiferromagnets with weak ferromagnetism, in which the Dzyaloshinskii--Moriya interaction causes canting of the magnetic sublattices and the appearance of a small net magnetization. These materials possess quasi-ferromagnetic and quasi-antiferromagnetic spin modes \cite{balbashov1995}. 

The development of terahertz spectroscopy has made it possible to directly excite and study the spin modes of orthoferrites \cite{lu2012,kim2014apl,amelin2018}. Experimental studies of strong magnon-photon coupling in orthoferrites are summarized in the review \cite{caoMaReview2019}, and the excitation of terahertz magnons is discussed in \cite{safin2020}. Terahertz magnon-polaritons were experimentally observed in $\mathrm{TmFeO_3}$ \cite{grishunin2018}; subsequently, strong coupling of magnons with phonon polaritons \cite{sivarajah2019}, ultrastrong coupling of magnon modes \cite{makihara2021}, and a controllable transition between the weak and strong coupling regimes in $\mathrm{YFeO_3}$ \cite{baydin2023} were investigated. A method for controlling the strength of plasmon-magnon interaction was also studied in Ref.~\cite{pikalov2021}.

The efficiency of the magnon-photon interaction is determined not only by the coupling constant $g$, but also by the lifetimes of the interacting modes. In microwave magnonics, the use of high-Q cavities is one of the main ways to achieve high cooperativity \cite{rameshti2022}. This approach has also been realized for antiferromagnets, but for a low-frequency spin mode: strong coupling was observed in hematite $\alpha$-$\mathrm{Fe_2O_3}$ placed in an external cavity with frequencies of 18--45 GHz \cite{boventer2023}.
\begin{figure*}[htb]
    \centering
    \includegraphics[width=0.35\textwidth]{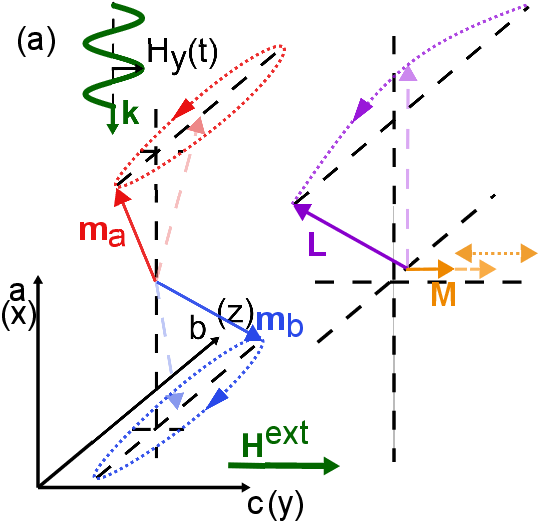}
    \hfill
    \includegraphics[width=0.4\textwidth]{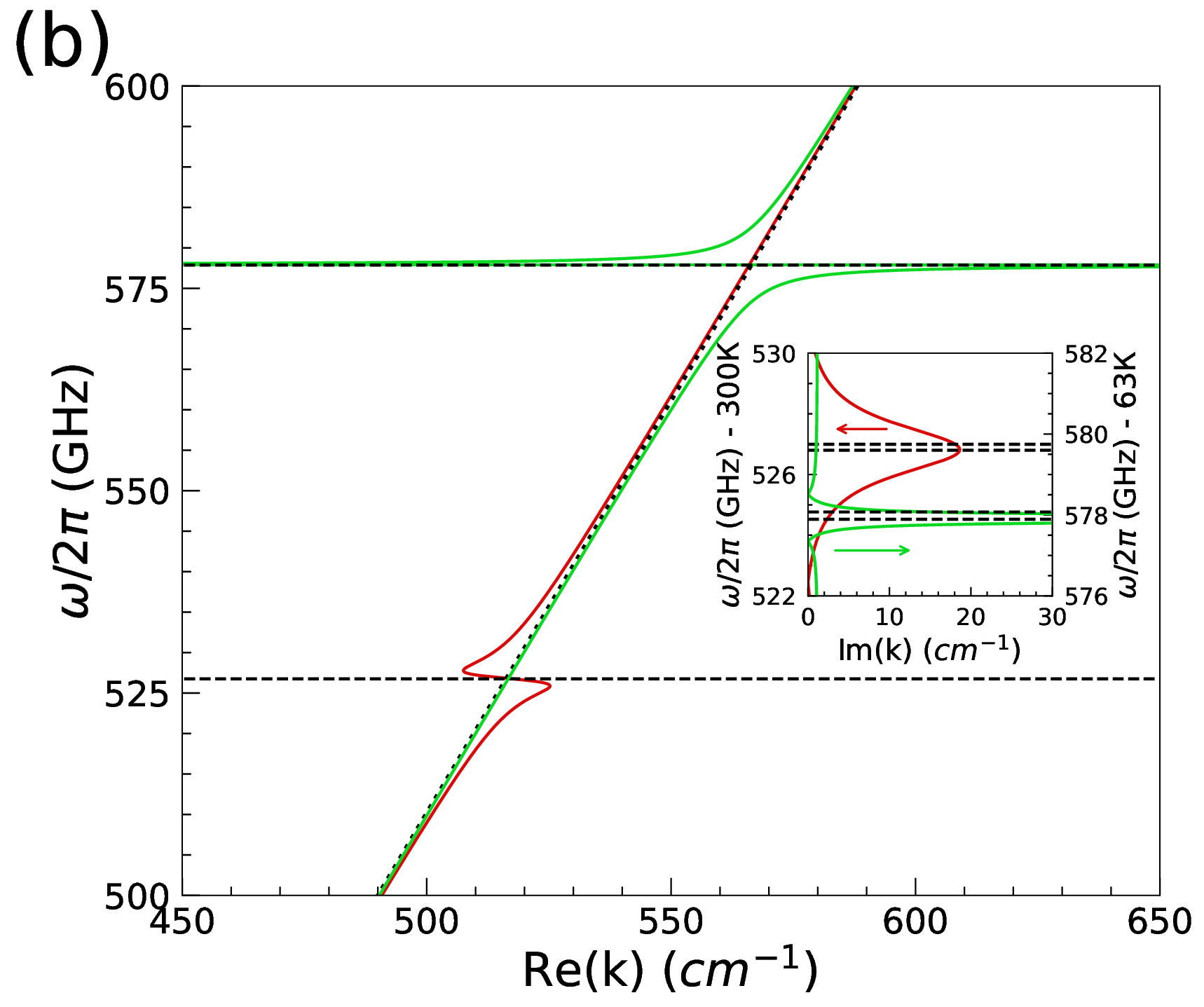}
  \caption{Quasi-antiferromagnetic magnon-polariton mode. (a) Trajectories of the sublattice magnetization vectors $\textbf{m}_a$ and $\textbf{m}_b$, as well as of the total magnetization $\textbf{M}$ and the N\'eel vector $\textbf{L}$. Dashed arrows show the equilibrium positions of the vectors. The figure is scaled so as to preserve the ratios between the angles of the spin dynamics. (b) Dispersion curves obtained from Eq.~(\ref{eq:AFM_dispersion}) at zero external field and at 300~K (red curve) and 63~K (green curve). Dashed lines show the dispersion of the noninteracting photon and magnon modes. The inset shows the dependence $\omega(\mathrm{Im}(k))$ near the forbidden gap. Black dashed lines in the inset show the boundaries of the forbidden gap.}
    \label{fig:AFM}
\end{figure*}
In the terahertz range, magnon-polaritons in antiferromagnets have been experimentally studied mainly using Fabry--Perot resonances formed by the magnetic slab \cite{grishunin2018}. In particular, strong coupling of the quasi-ferromagnetic and quasi-antiferromagnetic modes with such photonic resonances was investigated in $\mathrm{Gd_{0.5}Ho_{0.5}FeO_3}$ \cite{chen2025}. In addition, the interaction of antiferromagnetic resonances with electromagnetic modes was realized in a metallic tubular cavity \cite{bialek2021}. In Ref.~\cite{yavorskiy2025}, the use of an external tunable THz Fabry--Perot cavity formed by a liquid crystal and dielectric plates made it possible to vary the detuning between the photonic and magnon modes and to control the degree of their hybridization. However, the potential of specially designed high-Q THz cavities, for example photonic crystals, for increasing the cooperativity of the magnon-photon interaction remains little explored.

In this work, terahertz magnon-polaritons in a rare-earth orthoferrite are theoretically investigated, and for the first time the use of a high-Q THz cavity is proposed to achieve the regime of high cooperativity. Numerical calculations are performed for yttrium orthoferrite $\mathrm{YFeO_3}$, which is distinguished by high-Q spin resonances and unusual dynamic properties \cite{chetkin1980}. Both the intrinsic Fabry--Perot resonances of the magnetic slab and a photonic crystal cavity (PhCC) with Bragg mirrors are considered.

	\textbf{2. Lagrangian formalism and dynamical equations.}\\
	Consider a crystalline orthoferrite slab on which an electromagnetic wave is incident, with a frequency comparable to the spin resonance frequencies in the orthoferrite. We study the configuration in which the crystal axis $a$ is perpendicular to the slab plane, and the electromagnetic wave propagates along the normal to the slab, i.e., along the $a$ axis (Fig.~\ref{fig:AFM}a). We use a Cartesian coordinate system with $x\parallel a$, $y\parallel c$, $z\parallel b$; thus $\mathbf{k}\parallel x\parallel a$ and $\mathbf{H}^{ext}\parallel y\parallel c$. The transversality condition applies to the magnetic induction, $B_x=0$, whereas the longitudinal component of the magnetic field $H_x$ may be nonzero.% In what follows, we neglect the back-action of $H_x$ on ... \cite{supplementary_A} and consider the transverse components $H_y$ and $H_z$.
    
    To describe the system as a whole, we use the dimensionless magnetization vector $\textbf{m} = (\textbf{m}_a + \textbf{m}_b)/2$ and the antiferromagnetism vector (N\'eel vector) $\textbf{l} = (\textbf{m}_a - \textbf{m}_b)/2$, where $\textbf{m}_a$ and $\textbf{m}_b$ are the unit magnetization vectors of the two sublattices \cite[Appendix A]{supplementary_A}. Here $M_0$ is the sum of the magnitudes of the magnetizations of both sublattices per crystal unit cell: $\mathbf{M}_A=(M_0/2)\mathbf{m}_a$, $\mathbf{M}_B=(M_0/2)\mathbf{m}_b$, so that the total magnetization is $\mathbf{M}=\mathbf{M}_A+\mathbf{M}_B=M_0\mathbf{m}$. The vectors $\mathbf{m}$ and $\mathbf{l}$ satisfy the conditions $\mathbf{m}\cdot\mathbf{l}=0$ and $\mathbf{m}^2+\mathbf{l}^2=1$. It is convenient to describe the dynamics of these vectors in spherical coordinates, specifying the N\'eel vector $\textbf{l}$ by the angles $\theta$ and $\phi$ and introducing the canting angles of the sublattice magnetizations $\beta$ and $\delta$ along the azimuthal and polar angles, respectively, which remain small even for large deviations of $\textbf{l}$ from equilibrium: $\delta\ll1,\ \beta\ll1$.\\
	The equilibrium state and the spin dynamics of the magnet can be described within the Lagrangian formalism \cite{belov1979, zvezdin1979, andreev1980}, using the Lagrangian density $L=T-U_{mag}+L_{em}$, where $T$ is the kinematic term, $U_{mag}$ is the potential energy density of the magnetic system, and $L_{em}$ is the contribution of the electromagnetic field. All contributions to $L$ and $U_{mag}$ are referred to unit volume. 
    
    The potential energy density of the magnetic system consists of several contributions:
	\begin{equation}
		U_{mag} = U_{ex} + U_{ex}^l + U_{d} + U_Z + U_{a}.
		\label{eq:U_pot_all}
	\end{equation}
    The main contribution to the potential energy comes from the intersublattice exchange interaction: $U_{ex}=\lambda_{ex}M_0^2\mathbf{m_a} \cdot \mathbf{m_b}$, where $\lambda_{ex}$ is the intersublattice exchange constant. The intrasublattice interaction $U_{ex}^{l}$ determines the spatial distribution of the magnetization and gives rise to exchange spin waves: $U_{ex}^{l} =  A(\nabla \textbf{l})^2$, where $A$ is the inhomogeneous exchange constant. The Dzyaloshinskii--Moriya exchange interaction determines the static canting of the sublattices: $U_d= d_1l_x m_y+d_2l_y m_x$, where $d_1$ and $d_2$ are the Dzyaloshinskii constants, which in the case of yttrium orthoferrite can be considered equal in magnitude: $d_1\approx -d_2=d$. Orthoferrites have rhombic magnetic anisotropy, so the anisotropy energy contains two contributions with respect to the basis planes $ac$ and $ab$, with the corresponding constants $K_b$, $K_c$: $U_a= K_b l_z^2 - K_c l_x^2$. The action of the magnetic field of the electromagnetic wave on the magnetic sublattices is described by the Zeeman term: $U_Z = -M_0\textbf{m}\cdot(\textbf{H} + \textbf{H}^{ext})$, where $\textbf{H}$ is the magnetic field of the THz radiation and $\textbf{H}^{ext}$ is a static external magnetic field directed along the $c(y)$ axis. In what follows, we express all contributions to the potential energy in terms of the angular variables \cite[Appendix A]{supplementary_A}.

    Minimization of the potential energy density $U_{mag}$ for yttrium orthoferrite in the absence of the incident electromagnetic wave gives the equilibrium positions of the vectors $\mathbf{m}$ and $\mathbf{l}$: $\theta_0 =\pi/2,\, \varphi_0 = 0 \text{ or}\  \pi ,~\delta=0,\ \beta_0 = -d/(4\lambda_{ex} M_0^2) \pm \gamma H^{ext}/(4\lambda_{ex} M_0 \gamma)$; in the absence of an external field, $\beta_0\approx -0.0145$~rad. 
    Thus, in equilibrium the N\'eel vector is directed along the crystallographic $a$ axis (along the $x$ axis), and the canting of the sublattices leads to a small static magnetization along the crystallographic $c$ axis (along the $y$ axis).

    The magnetic field of the incident electromagnetic wave excites spin dynamics, as a result of which the angles describing the vectors $\mathbf{m}$ and $\mathbf{l}$ become time-dependent. In what follows, it is convenient to work with the deviation angles from the equilibrium position: $\theta_1=\theta-\theta_0$, $\varphi_1=\varphi-\varphi_0$, $\beta_1=\beta-\beta_0$, and $\delta_1=\delta-\delta_0$. These angles lead to the following dynamic components of the magnetization and N\'eel vectors: $m_x=\mp\beta_0\varphi_1, m_y=\pm \beta_1, m_z= \delta_1, l_x= \pm1, l_y= \pm\varphi_1, l_z= -\theta_1$, where the choice of the sign $\pm$ is determined by the equilibrium position $\varphi_0 =0$ and $\pi$, respectively. Here $m_x$, $m_y$, $m_z$ denote the dynamic components; the equilibrium component is $m_{y0}=\pm\beta_0$. \\
    We consider linear excitations, i.e., we assume that all deviation angles are much smaller than unity. Then, omitting constant terms, the Lagrangian density of the system is written as:
\begin{align}
    &L = \frac{M_0}{\gamma}(\delta_1\dot{\varphi_1} + \beta_1\dot{\theta_1}) - \frac{M_0}{\gamma}[\omega_{ex}\frac{\delta_1^2 + \beta_1^2}{2} + \omega_{ex}\beta_1\beta_0 - \notag\\ &\omega_{ex}\beta_0^2\frac{\theta_1^2}{2} + \omega_b \frac{\theta_1^2}{2} + \omega_{c0} \frac{\theta_1^2 + \varphi_1^2}{2} +\omega_D\beta_1 - \omega_D\beta_0\theta_1^2]\mp \notag\\ &\beta_0\varphi_1 M_0H_x\pm\beta_1 M_0 H_y + \delta_1 M_0 H_z \pm \beta_1 M_0 H^{ext} \notag\\
    &\mp \beta_0 M_0 \frac{\theta_1^2 + \varphi_1^2}{2}H^{ext} -A((\nabla\theta_1)^2 + (\nabla \varphi_1)^2) + \frac{1}{8\pi}(\hat{\varepsilon}\textbf{E}^2 - \textbf{B}^2),
    \label{eq:Lagrangian}
\end{align}
where the first term is the kinematic term, $\gamma$ is the gyromagnetic ratio, $\omega_b = 2K_b\gamma/M_0$, $\omega_{c0} = 2K_c\gamma/M_0$, $\omega_{ex} = 4\lambda_{ex} M_0 \gamma$, $\omega_D = \gamma d / M_0$, $\hat{\varepsilon}=\mathrm{diag}(\varepsilon_x,\varepsilon_y,\varepsilon_z)$ is the dielectric permittivity tensor of the medium, and $\textbf{E}$ and $\textbf{B}$ are the electric field and magnetic induction vectors, respectively. The last term in Eq.~(\ref{eq:Lagrangian}) defines the contribution of the electromagnetic field to the Lagrangian. 

%In the adopted coordinate system at room temperature $\varepsilon_z=25.4+0.04i$ and $\varepsilon_y=21.9+0.09i$ \cite{balbashov1995}. 

Using the Euler--Lagrange equations, we obtain expressions describing the dynamics of the magnetic sublattices and of the electromagnetic field in the orthoferrite. We take dissipation of the magnetic system into account phenomenologically by introducing damping coefficients $\kappa_{m1}$ and $\kappa_{m2}$ for spin oscillations excited by the electromagnetic field components $H_z$ and $H_y$, respectively. Thus, we obtain:
\begin{align}
&\ddot{\theta}_1+\kappa_{m2}\dot{\theta}_1+
\omega_{r2}^2\theta_1-c_m^2\partial_x^2\theta_1
=\mp\gamma\dot{H}_y,\notag\\
&\ddot{\beta}_1+\kappa_{m2}\dot{\beta}_1+
\omega_{r2}^2\beta_1-c_m^2\partial_x^2\beta_1
=
\pm\gamma\omega_aH_y \mp\frac{c_m^2}{\omega_{ex}}\gamma\partial_x^2H_y,\notag\\
&\ddot{\varphi}_1+\kappa_{m1}\dot{\varphi}_1+
\omega_{r1}^2\varphi_1-c_m^2\partial_x^2\varphi_1
=-\gamma\dot{H}_z\mp\beta_0\gamma\omega_{ex}H_x,\notag\\
&\ddot{\delta}_1+\kappa_{m1}\dot{\delta}_1+
\omega_{r1}^2\delta_1-c_m^2\partial_x^2\delta_1
=
\gamma\omega_cH_z-\frac{c_m^2}{\omega_{ex}}\gamma\partial_x^2H_z\mp\beta_0\gamma\dot{H_x},\notag\\
&H_x + 4\pi M_0m_x = 0\notag\\
&\partial_x^2 H_y
-\frac{\varepsilon_z}{c^2}\ddot H_y
=
\frac{4\pi M_0\varepsilon_z}{c^2}\ddot m_y,\notag\\
&\partial_x^2H_z
-\frac{\varepsilon_y}{c^2}\ddot H_z
=
\frac{4\pi M_0\varepsilon_y}{c^2}\ddot m_z.
\label{eq:dynamic}
\end{align}
where $c$ is the speed of light in vacuum, $c_m =\sqrt{2\gamma A\omega_{ex}/M_0}$ is the spin-wave velocity in the antiferromagnet, $\omega_{r1}^2 = \omega_{ex}\omega_c$, $\omega_{r2}^2 = \omega_{ex}\omega_a$, $\omega_c = \omega_{c0} + \gamma H^{ext}/ \omega_{ex} \cdot(\omega_D + \gamma  H^{ext})$, $\omega_a = \omega_b + \omega_{c0} +\omega_D/\omega_{ex}\cdot(\omega_D + \gamma  H^{ext})$, and $\kappa_{m1,m2}$ are the damping coefficients of the magnetic modes. For $\kappa_{mi}\ll\omega_{ri}$, the damping coefficients are equal to the full width at half maximum (FWHM) of the corresponding resonance line: $\Delta\omega_{mi}^{\rm FWHM}=\kappa_{mi}$. \\

\textbf{3. Magnon-polariton modes}\\
	From Eq.~(\ref{eq:dynamic}) it follows that two magnon-polariton modes are excited in this configuration. We seek solutions in the form $\{\theta_1,\beta_1,\varphi_1,\delta_1,\mathbf{H}\}\propto\exp[i(kx-\omega t)]$.
    In one of them, together with the wave field component $H_y$, the polar angle $\theta_1$ and the azimuthal canting angle $\beta_1$ oscillate, while the remaining electromagnetic field components ($H_z$, $H_x$) and angles ($\varphi_1$, $\delta_1$) are zero:
\begin{align}
    \beta_1 &=\pm\frac{\gamma H_y\left(\omega_a+c_m^2k^2/\omega_{ex}\right)}{\omega_{\rm m2}^2(k)-\omega^2-i\kappa_{m2}\omega},\notag\\[2mm]
\theta_1 &=\pm\frac{i\gamma\omega H_y}{\omega_{\rm m2}^2(k)-\omega^2-i\kappa_{m2}\omega},
\label{eq:Hy_mode_main}
\end{align}
where $\omega_{\rm m2}^2(k)=\omega_{r2}^2+c_m^2k^2$ is the dispersion of the magnon mode noninteracting with photons.
    The vector $\textbf{L}$ oscillates in the $ab$ plane, while the magnetization vector remains directed along the $c$ axis but changes in magnitude (Fig.~\ref{fig:AFM}(a)). Therefore, this spin mode is usually called quasi-antiferromagnetic (AFM). 
\begin{figure*}[t]
    \centering
        \includegraphics[width=0.4\textwidth]{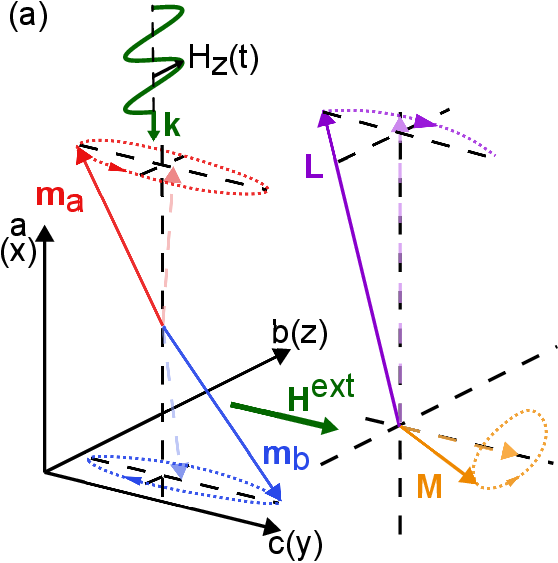}
    \hfill
    \includegraphics[width=0.4\textwidth]{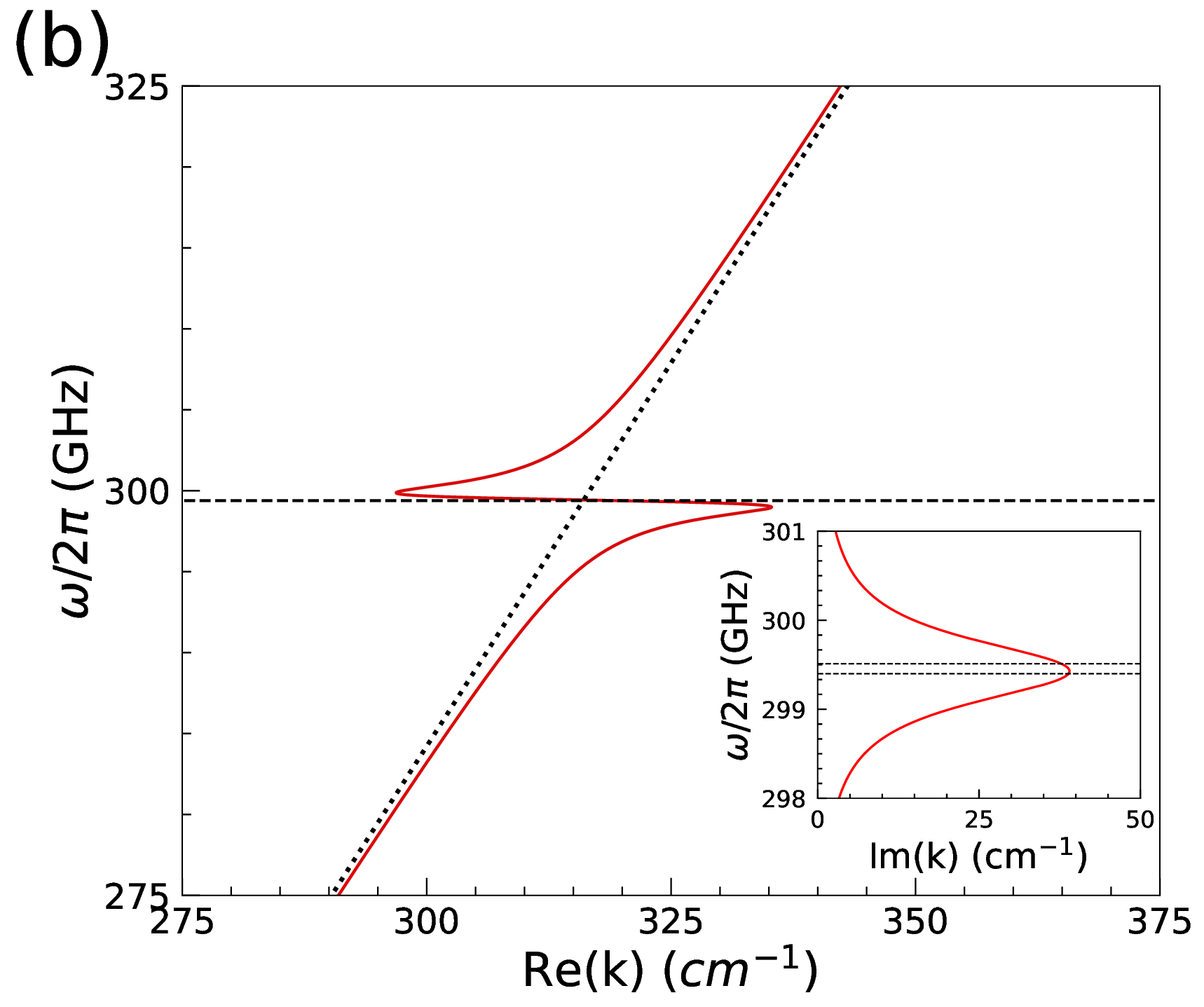}
  \caption{Quasi-ferromagnetic magnon-polariton mode. (a) Trajectories of the sublattice magnetization vectors $\textbf{m}_a$ and $\textbf{m}_b$, as well as of the total magnetization $\textbf{M}$ and the N\'eel vector $\textbf{L}$. Dashed arrows show the equilibrium positions of the vectors. The figure is scaled so as to preserve the ratios between the angles of the spin dynamics. (b) Dispersion of the magnon-polariton obtained from Eq.~(\ref{eq:FM_dispersion}) at 300~K. Dashed lines show the dispersion of the noninteracting photon and magnon modes. The inset shows the dependence $\omega(\mathrm{Im}(k))$ near the forbidden gap. Black dashed lines in the inset show the boundaries of the forbidden gap.}
   \label{fig:FM}
\end{figure*}
From the condition for the existence of a nontrivial solution of system (\ref{eq:dynamic}), we obtain the dispersion relation for the AFM magnon-polariton:
\begin{align}
    \left[
\omega_{\rm ph}^2(k)-\omega^2
\right]
\left[
\omega_{\rm m2}^2(k)-\omega^2-i\kappa_{m2}\omega
\right]
=\notag\\
\omega^2\omega_M
\left(
\omega_a+c_m^2k^2/\omega_{ex}
\right),
\label{eq:AFM_dispersion}
\end{align}

where $\omega_{\rm ph}(k)=ck/\sqrt{\varepsilon_z}$ is the dispersion of the photon mode and $\omega_M=4\pi M_0\gamma$.
For the parameters of $\mathrm{YFeO_3}$ at room temperature \cite[Appendix B]{supplementary_A}, $\omega_{r2}/2\pi=526.8$~GHz and $\kappa_{m2}/2\pi=2.1$~GHz, i.e., $\kappa_{m2}/2\ll\omega_{r2}$. %The coefficient $\kappa_{m2}$ characterizes the damping of the magnetic mode and corresponds to the full width at half maximum (FWHM) of the magnetic resonance.

 At the frequency of the AFM resonance, the wavelength of terahertz radiation in the orthoferrite is about 120~$\mu$m, which significantly exceeds the crystal lattice period and makes it possible to neglect the influence of intrasublattice exchange. Neglecting dissipation, from Eq.~(\ref{eq:AFM_dispersion}) we obtain a quadratic equation in $\omega^2$, which determines two magnon-polariton branches:

\begin{equation}
\begin{aligned}
&\omega_{\pm}^2(k)
=\frac{1}{2}
\left[
\omega_{\rm ph}^2(k)+\omega_{\rm r2}^2
+\omega_M\omega_a
\right]\\
&\pm\frac{1}{2}
\Bigg\{
\left[
\omega_{\rm ph}^2(k)+\omega_{\rm r2}^2
+\omega_M\omega_a
\right]^2
-4\omega_{\rm ph}^2(k)\omega_{\rm r2}^2
\Bigg\}^{1/2}
\end{aligned}
\label{eq:AFM_polariton_branches}
\end{equation}

Far from the resonance region, the solutions $\omega_{\pm}(k)$ correspond predominantly to the magnon and photon modes, whereas near their intersection hybridization occurs, leading to the characteristic anticrossing of the magnon-polariton branches.

	%where $\Delta_2 = \omega_{r2}^2-\omega^2+i\kappa_{m2}\omega$, $\kappa_{m2}$ is the width (FWHM) of the quasi-antiferromagnetic resonance, $\omega_{r2}^2=\omega_{ex}(\omega_b+\omega_c)=\omega_{ex}\omega_aH^{ext}$. For the parameters of $\mathrm{YFeO_3}$ at room temperature (see Appendix A) $\omega_{r2}/2\pi=526.8$~GHz. 

    We note that the dispersion relation of the magnon-polariton can also be represented in the form $k = k_0 \sqrt{\varepsilon_z\mu_{yy}}$, where the diagonal component of the magnetic permeability tensor is $\mu_{yy}=1+\Delta\mu_2\omega_{r2}^2/(\omega_{r2}^2-\omega^2-i\omega\kappa_{m2})$, $k_0=\omega/c$ is the wave number in vacuum, and $\Delta\mu_2=\omega_M/\omega_{ex}$ is the static contribution of the AFM resonance to the magnetic permeability.
    
    A forbidden gap arises near the resonance frequency: $\Delta\omega_{\mathrm{gap}}/2\pi =\omega_M\omega_a/(4\pi\omega_{r2})\simeq0.2$~GHz. At low temperatures (e.g., 63~K), the width of the AFM resonance ($\kappa_{m2}/2\pi=50$~MHz) \cite{balbashov1995} becomes smaller than the width of the polariton forbidden gap, which may allow the latter to be observed experimentally (Fig.~\ref{fig:AFM}b).

Near the AFM resonance, Eq.~(\ref{eq:AFM_polariton_branches}) takes the standard form of the polariton splitting:
\begin{equation}
\omega_{\pm}=\frac{1}{2}\left(\omega_{\mathrm{ph}}+\omega_{r2}
\pm\sqrt{(\omega_{\mathrm{ph}}-\omega_{r2})^2+4g^2}\right).
\end{equation}
 Here $g=\sqrt{\omega_M\omega_a}/2$ determines half of the resonant splitting of the two hybrid polariton branches and represents the magnon-photon coupling constant. Equivalently, in terms of the magnetic permeability parameters, $g=\omega_{r2}\sqrt{\Delta\mu_2}/2$. For the AFM mode of $\mathrm{YFeO_3}$, we obtain $g/2\pi\simeq7$~GHz.

 Let us introduce the effective mass
$m_{\rm eff}^{(\pm)}=
\hbar
\left(
\partial^2\omega_{\pm}/{\partial k^2}
\right)^{-1}$, which characterizes the dispersion properties of the magnon-polariton and the change of its group velocity with momentum. Near the crossing point of the noninteracting magnon and photon modes, from Eq.~(\ref{eq:AFM_polariton_branches}) we obtain
\begin{equation}
m_{\rm eff}^{(\pm)}
\simeq
\pm\frac{4\hbar g\varepsilon_z}{c^2}
=
\pm\frac{2\hbar\varepsilon_z}{c^2}
\sqrt{\omega_M\omega_a}.
\label{eq:effective_mass_res}
\end{equation}
Thus, the effective masses of the upper and lower magnon-polariton branches at the center of the anticrossing have equal magnitudes and opposite signs. For $\mathrm{YFeO_3}$ we obtain
$|m_{\rm eff}|\simeq5\times10^{-39}$~kg$=5\times10^{-9}m_e$, where $m_e$ is the electron mass. The small effective mass of the magnon-polariton corresponds to a strong dynamic response to external perturbations. 

Along with the AFM mode, excitation of the quasi-ferromagnetic (FM) mode is also possible. In the configuration considered, it hybridizes with the component $H_z$ of the magnetic field of the electromagnetic wave. In this mode, together with the magnetic field of the THz wave $H_z$, the angles $\varphi_1$ and $\delta_1$ oscillate, while the transverse field component $H_y$ and the angles of the spin dynamics ($\theta_1$, $\beta_1$) are zero (Fig.~\ref{fig:FM}(a)):
	\begin{equation}
	\label{eq:Hz_mode_main}
		\begin{aligned}
\delta_1 &= \frac{H_z \gamma (\omega_c + c_m^2k^2/\omega_{ex})\pm i\beta_0\gamma\omega H_x}{\omega_{m1}^2(k) - \omega^2 - i\kappa_{m1}\omega}\\
			\varphi_1 &= \frac{i\omega \gamma H_z \mp \beta_0\gamma\omega_{ex}H_x}{\omega_{m1}^2(k) - \omega^2 - i\kappa_{m1}\omega}
		\end{aligned}
	\end{equation}
    where $\omega_{m1}^2(k) = \omega_{r1}^2 + c_m^2k^2$. The contribution of $H_x$ to the dispersion is neglected, since it gives corrections of the order of $\beta_0^2 \approx 0.0002$.
	Similarly to the previous mode, we obtain the dispersion relation for the FM mode (Fig.~\ref{fig:FM}(b)):
    \begin{equation}
\begin{aligned}
\left[
\omega_{\rm ph, F}^2(k)-\omega^2
\right]
\left[
\omega_{\rm m1}^2(k)-\omega^2-i\kappa_{m1}\omega
\right]
=\\
\omega^2\omega_M
\left(
\omega_c+c_m^2k^2/\omega_{ex}
\right),
\label{eq:FM_dispersion}
\end{aligned}
\end{equation}
where $\omega_{\rm ph, F}(k)=ck/\sqrt{\varepsilon_y}$ is the dispersion of the photon mode. Near the resonance, the coupling constant of the FM mode is $g_{\rm FM}=\sqrt{\omega_M\omega_c}/2$, which for the parameters of $\mathrm{YFeO_3}$ at room temperature gives $g_{\rm FM}/2\pi\simeq4.0$~GHz.\\ %$\omega_{r1}/2\pi=299.4$~GHz. $\kappa_{m1} / 2\pi = 0.9$~GHz is the width (FWHM) of the quasi-ferromagnetic resonance,

\textbf{4. Magnon-polaritons in an orthoferrite slab}\\

\begin{figure*}[t]
    \centering
        \includegraphics[width=0.8\textwidth]{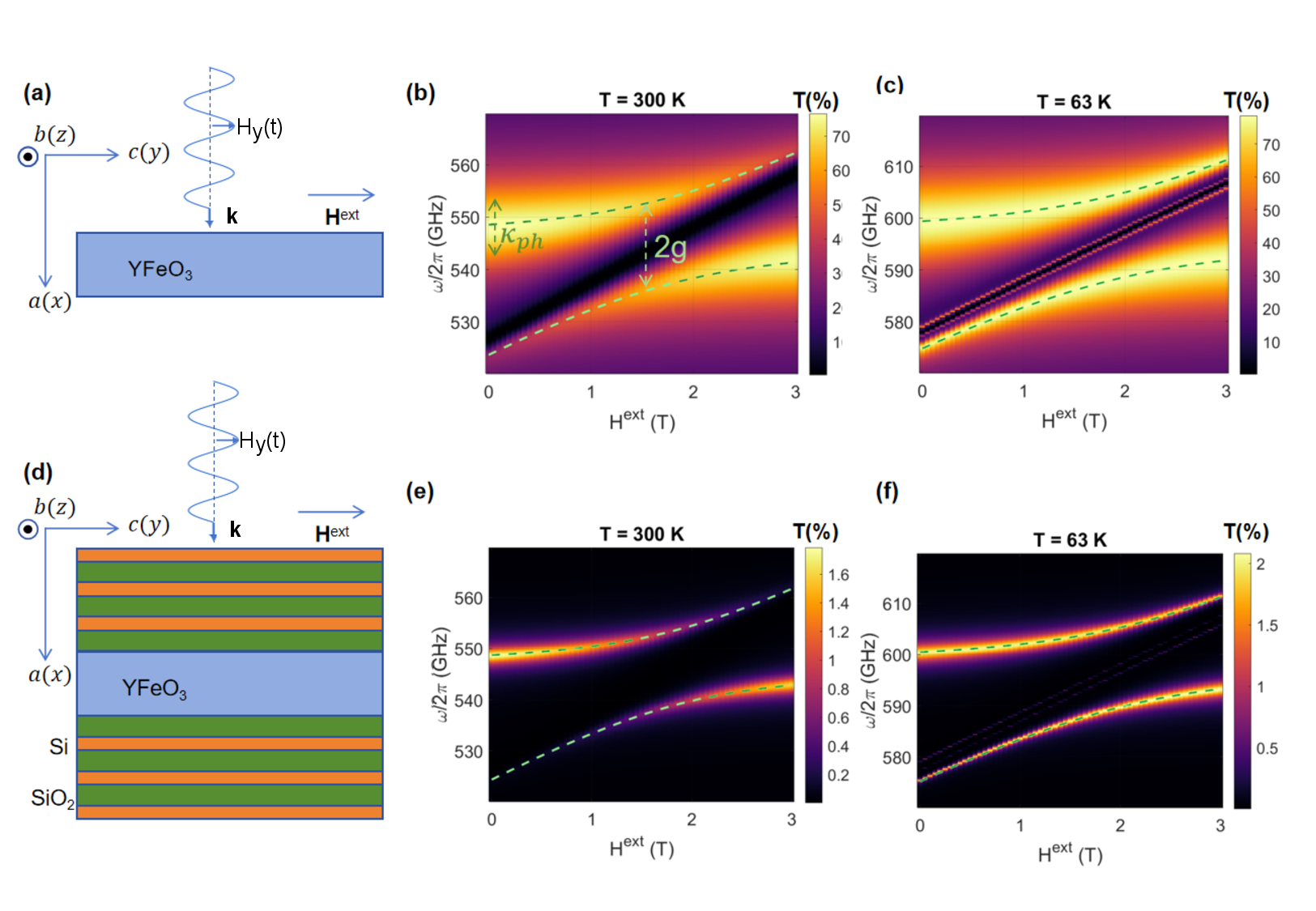}
 \caption{Magnon-polaritons upon hybridization of the AFM resonance and the Fabry--Perot resonance in a free yttrium orthoferrite slab (a--c) and in a slab inside a PhCC (d--f). (a,d) Schematics of the structure and the incident radiation. (b,c,e,f) Calculated transmission spectra of THz radiation at frequencies near the excitation of the AFM mode: for the slab at temperatures of 300~K (b) and 63~K (c), and for the photonic crystal at temperatures of 300~K (e) and 63~K (f). The PhCC is formed by two Bragg mirrors, each consisting of alternating $\mathrm{Si/SiO_2}$ layers. The calculation was performed for 3 pairs of layers on each side. The layer thicknesses are given in Ref.~\cite[Appendix C]{supplementary_A}. Dashed lines show the fit of the resonances by Eq.~(7).}
   \label{fig:FP}
\end{figure*}

In experimental studies of antiferromagnets, crystalline slabs are usually used. Owing to the large refractive index in the terahertz range, such a slab itself forms a Fabry--Perot (FP) cavity (Fig.~\ref{fig:FP}(a)). For the AFM mode considered, the width of the photonic resonance is $\kappa_{\mathrm{ph}}/2\pi=20$~GHz, which corresponds to $Q=\omega_{\mathrm{ph}}/\kappa_{\mathrm{ph}}\simeq 26$ at a frequency of $527$~GHz. Despite the moderate quality factor, when the AFM resonance is tuned by an external magnetic field, a pronounced anticrossing of the photonic and magnon branches appears in the transmission spectrum (Fig.~\ref{fig:FP}(b)). The spectral positions of the upper and lower polariton branches can be fitted by Eq.~(7), and the value of $g$ can be determined from the fit.

For a quantitative characterization of the interaction, we further use the cooperativity
\begin{equation}
C=\frac{4 g^2}{\kappa_{\mathrm{ph}}\kappa_m}.
\end{equation}
The cooperativity compares coherent hybridization with dissipative losses; large values of $C$ are especially important for coherent energy transfer and conversion, information storage, and quantum magnonics. 
For the intrinsic FP cavity of the slab at room temperature, we obtain $C\simeq6$ for the AFM mode and $C\simeq 4$ for the FM mode. Upon cooling to 63~K, the magnon line of the AFM mode sharply narrows to $\kappa_{m2}/2\pi=0.05$~GHz \cite{balbashov1995}, which leads to an increase of the cooperativity to $C\simeq260$. In this case, as follows from Fig.~\ref{fig:FP}c (at $H^{\mathrm{ext}}=1.5$~T), the width of the magnon-polariton resonance is $\kappa_{mp}/2 \pi= 13.5$~GHz, which corresponds to an energy lifetime of the magnon-polariton $\tau_E=1/\kappa_{mp}= 12$~ps.
 At low temperatures, additional narrow resonances appear in the transmission spectra near the anticrossing (Fig.~\ref{fig:FP}(c)). The calculation shows that, for a small value of $\kappa_{m}$, the wave vector $k$ of the AFM mode can satisfy several Fabry--Perot resonance conditions ($k = \pi m/d$) within a small frequency interval near the resonance. Therefore, besides the main anticrossing, side branches from lower- and higher-order resonances are also observed. Thus, in Fig.~\ref{fig:FP}(c), the main line in the spectrum corresponds to the Fabry--Perot resonance with $m=8$, while the two narrow transmission resonances are due to the hybridization of magnons with the Fabry--Perot modes with $m=7$ and 9. \\
 
\textbf{5. Magnon-polaritons in a photonic crystal}\\
A further increase in the cooperativity is possible by using higher-Q cavities. Let us consider a yttrium orthoferrite slab inside a photonic crystal cavity (PhCC) formed by two Bragg mirrors consisting of alternating pairs of layers with different dielectric permittivities $\varepsilon_{d,i}$, $i= $ 1,2 (Fig.~\ref{fig:FP}(d)). In this work, silicon and silicon dioxide are chosen as the materials of such a structure \cite{chaudhary2024}. The transmission spectra are calculated by the transfer matrix method \cite{yeh1980},\cite[Appendix D]{supplementary_A}.

\setcounter{table}{0}
\begin{table*}[h!]
    \centering
    \caption{Parameters of magnon-polaritons in a PhCC containing yttrium orthoferrite.}
    \label{tab:params}
    \begin{tabular}{|c|c|c|c|c|c|c|c|}
    \hline
    & $g/2\pi$, GHz & $\kappa_{m,i}/2\pi$, GHz & $\kappa_{ph}/2\pi$, GHz & $C$ & $\kappa_{mp}/2\pi$, GHz & $\tau_E$, ns \\
    \hline
    AFM, $H_{\mathrm{THz}}\parallel c$, 300 K & 7.3 & 2.1 & 2.5 & 41 & 2.30 & 0.07 \\
    \hline
    AFM, $H_{\mathrm{THz}}\parallel c$, 63 K & 7.9 & 0.05 & 2.2 & 2200 & 1.15 & 0.14 \\
    \hline
    FM, $H_{\mathrm{THz}}\parallel b$, 300 K & 4.1 & 0.90 & 1.4 & 53 & 1.2 & 0.14 \\
    \hline
    \end{tabular}
\end{table*}
As a rule, to obtain a high-Q FP resonance, the layers of the Bragg mirrors are made quarter-wave: $d_i=\lambda_0/4\sqrt{\varepsilon_{d,i}}$, and the thickness of the resonator layer is $d=\lambda_0 m/2\sqrt{\varepsilon}$, where $\lambda_0$ is the wavelength of light in vacuum for which the PhCC is designed and $m$ is an integer (Fig.~\ref{fig:FP}(e)). In this case, the FP mode lies at the center of the photonic band gap of the PhCC \cite[Fig.~S1]{supplementary_A}.

For the cavity considered, the use of three pairs of layers in each Bragg mirror yields a photonic band gap 300~GHz wide and significantly increases the quality factor of the photonic mode: $\kappa_{\rm ph}/2\pi = 2.5$~GHz for the AFM mode and $\kappa_{\rm ph}/2\pi = 1.4$~GHz for the FM mode. As a result, the magnon-photon anticrossing becomes more pronounced (Fig.~\ref{fig:FP}(e)), and the cooperativity increases to $C=41$ for the AFM mode and $C=red$ for the FM mode.   

Upon cooling to 63~K, the AFM resonance narrows significantly and, in addition, its frequency increases by 51~GHz. As a result, the resonances do not cross in the range of fields considered, and to obtain their anticrossing a magnetic field of about 5~T must be applied opposite to the static magnetization \cite[Appendix C, Fig.~S2]{supplementary_A}. Therefore, for operation at low temperatures it is reasonable to use a PhCC with modified parameters, tuned to the new frequency.

As a result, the energy lifetime and the cooperativity of the magnon-polariton reach $\tau_E\sim 0.14$~ns and $C=2.2 \times 10^3$, respectively (Table~\ref{tab:params}, Fig.~\ref{fig:FP}(f)). At the same time, cooling has little effect on the properties of the FM mode and the cooperativity of the magnon-polariton associated with it.

We note that such high values of the cooperativity require a narrow AFM resonance line, i.e., single-crystal slabs, and the permissible deviation of the slab thickness from the design value (due to roughness or non-parallelism of the faces) is $\Delta d\lesssim1~\mu$m: in this case $\kappa_{ph}$ increases by no more than a factor of $\sqrt{2}$, while the splitting practically does not change \cite[Appendix E]{supplementary_A}.

\textbf{6. Conclusion}\\

In this work, terahertz magnon-polaritons in antiferromagnets with weak ferromagnetism have been theoretically investigated. Within the two-sublattice model, dispersion relations for the quasi-antiferromagnetic and quasi-ferromagnetic modes have been obtained; calculations have been performed for yttrium orthoferrite $\mathrm{YFeO_3}$. The resonant magnon-photon interaction forms two hybrid branches with a characteristic anticrossing. For the quasi-antiferromagnetic mode, $g/2\pi\simeq7$~GHz and an effective mass of about $5\times10^{-9}m_e$ have been obtained. The intrinsic Fabry--Perot resonances of the magnetic slab ensure the formation of magnon-polaritons; however, their moderate quality factor limits the cooperativity to $C\simeq6$ at room temperature.

Placing the magnetic slab in a photonic crystal cavity with Bragg mirrors reduces the width of the photonic resonance almost without changing $g$ and increases the cooperativity to $C=41$ at 300~K. At 63~K, in a cavity retuned to the new frequency of the AFM resonance, owing to the reduction of the magnon damping it increases to $C\simeq2.2\times10^3$. Thus, high cooperativity is achieved by controlling the lifetimes of the interacting modes. The combination of terahertz frequencies, small effective mass, and high cooperativity makes such states promising for coherent control of spin excitations, terahertz magnonics, and quantum technologies.
	
	\textbf{Funding.}
The theoretical and analytical part of the work was supported by the Russian Science Foundation (grant no. 23-62-10024). A.A.Luzgina, A.A.Kraev, and V.I.Belotelov thank the BASIS Foundation for the Development of Theoretical Physics and Mathematics (grant no. 25-1-1-49-1) for financial support of the modeling of magnon-polaritons in photonic crystals.
    
    \textbf{Conflict of interest.}
The authors of this work declare that they have no conflict of interest.

\end{document}

\maketitle
\pagestyle{empty}
\thispagestyle{empty}
\textbf{A. Derivation of the main equations}\\
Each crystal cell of $\mathrm{YFeO_3}$ contains 4 iron ions ($\mathrm{Fe^{3+}}$), the magnetization vectors of which are oppositely directed in pairs. Thus, we can speak of the presence of two equivalent sublattices (A and B) with the magnetizations $\mathbf{M}_A=(M_0/2)\mathbf{m}_a$ and $\mathbf{M}_B=(M_0/2)\mathbf{m}_b$, where $M_0$ is the sum of the magnitudes of the magnetizations of both sublattices per unit crystal cell, $\mathbf{m}_{a,b}$ are unit vectors of the magnetizations of the two sublattices. Complete magnetization of the structure is $\mathbf{M} = \mathbf{M}_a + \mathbf{M}_b$. To describe the system we introduce magnetization and Neel vectors as follows: $\mathbf{m} = (\mathbf{m}_a + \mathbf{m}_b) / 2$ and $\mathbf{l} = (\mathbf{m}_a - \mathbf{m}_b) / 2$. Note that in orthoferrites, the antiferromagnetic (AFM) ordering is not perfectly antiparallel; a slight deviation occurs, primarily due to Dzyaloshinskii--Moriya interaction. We use the Cartesian axes $x\parallel a$, $y\parallel c$, and $z\parallel b$, as in the main text. Vectors $\mathbf{m}$, $\mathbf{l}$ satisfy the conditions of the following equations $\mathbf{m}\cdot\mathbf{l} = 0$ and $\mathbf{m}^2 + \mathbf{l}^2 = 1$.

%Thus, magnetization and N\'eel vectors are defined as $\textbf{m} = (\textbf{m}_a + \textbf{m}_b)/{2}$ and $\textbf{l} = (\textbf{m}_a - \textbf{m}_b)/{2}$, accordingly. 
In the spherical coordinate system, vectors $\textbf{m}_a$, $\textbf{m}_b$ can be represented as: $\mathbf{m}_{\mathrm{a,b}} = (\sin \theta_{\mathrm{a,b}}\cos \varphi_{\mathrm{a,b}},~\sin \theta_{\mathrm{a,b}}\sin \varphi_{\mathrm{a,b}},~\cos \theta_{\mathrm{a,b}})$.

Let us define the N\'eel vector $\textbf{l}$ in the spherical coordinates by the angles $\varphi$ and $\theta$, while the skew (canting) angles  $\beta$ and $\delta$ will be used to describe a canting of the magnetizations of the two sublattices. Consequently,
\begin{equation}
		\begin{array}{r}
			\theta_{\mathrm{a}} = \theta - \delta ,\quad 
             \theta_{\mathrm{b}} = \pi - \theta - \delta,\\
             \varphi_{\mathrm{a}} = \varphi + \beta, \quad \varphi_{\mathrm{b}} = \pi + \varphi - \beta. 
		\end{array}
		\label{eq:angles}
	\end{equation}
Therefore, to first order in the canting angles, the components of the magnetization and the Néel vector are expressed in terms of $\theta$, $\varphi$, $\beta$, and $\delta$ as follows:
\begin{equation}
		\begin{aligned}
			m_x &= -\beta \sin \theta \sin \varphi -\delta \cos \theta \cos \varphi,\\
			m_y&= \beta \sin \theta \cos \varphi -\delta \cos \theta \sin \varphi,\\
			m_z&= \delta \sin \theta,\\
			l_x&= \sin \theta \cos \varphi,\\
			l_y&= \sin \theta \sin \varphi,\\
			l_z&= \cos \theta.
		\end{aligned}
		\label{eq:ml_full}
	\end{equation}
Thus, the following expressions for the magnetic energies and the kinetic term are obtained:
\begin{equation}
\begin{aligned}[c]
	U_{ex} &= \lambda_{ex}M_0^2\mathbf{m}_a \cdot \mathbf{m}_b
	\simeq -\lambda_{ex}M_0^2 
	+ 2\lambda_{ex}M_0^2 (\beta^2 \sin^2\theta + \delta^2), \\[4pt]
	U_{ex}^{l} &= A(\nabla \mathbf{l})^2
	= A\left[(\nabla\theta)^2 +
	\sin^2\theta(\nabla\varphi)^2\right], \\[4pt]
	U_d &= d_1l_xm_y+d_2l_ym_x
	= d\beta\sin^2\theta, \\[4pt]
	U_a &= K_bl_z^2 - K_cl_x^2
	= -K_c + K_b\cos^2\theta
	+ K_c(\sin^2\theta\sin^2\varphi + \cos^2\theta), \\[4pt]
	U_Z &= -M_0\mathbf{m}\cdot(\mathbf{H}+\mathbf{H}^{ext}) \\
	&= - M_0\,\left[(-\beta \sin \theta \sin \varphi -\delta \cos \theta \cos \varphi)H_x+(\beta\sin\theta\cos\varphi
	-\delta\cos\theta\sin\varphi)(H_y + H^{ext})
	+\delta\sin\theta H_z \right], \\[4pt]
	T &= \frac{M_0}{\gamma}\mathbf{m}\cdot
	(\mathbf{l}\times\dot{\mathbf{l}})
	= \frac{M_0}{\gamma}\sin\theta
	(\delta\dot{\varphi}+\beta\dot{\theta}).
\end{aligned}
\label{eq:energy}
\end{equation}
The electromagnetic contribution to the Lagrangian is
\[
L_{em}=\frac{1}{8\pi}
\left(\mathbf{E}\cdot\hat{\varepsilon}\mathbf{E}-\mathbf{B}^2\right).
\]
Here $U_{mag}=U_{ex}+U_{ex}^{l}+U_d+U_a+U_Z$, and constants $\lambda_{\text{ex}}, \, M_0, \, A, \, d_1 = -d_2 = d, \, K_{b, c}, \, \hat{\varepsilon}$ are introduced in B.\\ 
Using Eqs.~(\ref{eq:energy}), the full Lagrangian of the system is presented as $L=T-U_{mag}+L_{em}$:
\begin{equation}
    \begin{aligned}
        L &= \frac{M_0}{\gamma} \sin\theta
        [\beta \dot{\theta} + \delta \dot{\varphi}]
        - 2\lambda_{\text{ex}} M_0^2
        (\beta^2 \sin^2\theta + \delta^2)
        - A [(\nabla\theta)^2 + \sin^2\theta (\nabla\varphi)^2]
        - d \beta \sin^2\theta + K_c - K_b \cos^2\theta \\
        &\quad - K_c (\sin^2\theta \sin^2\varphi + \cos^2\theta)
        + M_0 [(-\beta \sin \theta \sin \varphi -\delta \cos \theta \cos \varphi)H_x+(\beta\sin\theta\cos\varphi
        - \delta\cos\theta\sin\varphi) H_y
        + \delta\sin\theta H_z \\
        &\qquad\quad + (\beta\sin\theta\cos\varphi
        - \delta\cos\theta\sin\varphi) H^{\text{ext}}]
        + \frac{1}{8\pi}
        (\mathbf{E}\cdot\hat{\varepsilon}\mathbf{E} - \mathbf{B}^2).
    \end{aligned}
    \label{eq:fullLagrangian}
\end{equation}
To find an equilibrium state, we set all spatial derivatives and projections of the incident electromagnetic wave to zero and then differentiate the magnetic potential energy with respect to each angle:
\begin{equation}
    \begin{aligned}
        \frac{\partial U_{mag}}{\partial \beta}&=\frac{M_0}{\gamma}\omega_{ex}\beta\sin^2\theta + \frac{M_0}{\gamma}\omega_D\sin^2\theta - M_0 \sin\theta\cos\varphi H^{ext} = 0\\
        \frac{\partial U_{mag}}{\partial \theta} &= \frac{M_0}{\gamma}\omega_{ex}\beta^2\sin\theta\cos\theta + 2\frac{M_0}{\gamma}\omega_D\beta\sin\theta\cos\theta - \frac{M_0}{\gamma}\omega_b\cos\theta\sin\theta + \frac{M_0}{\gamma}\omega_{c0}\sin\theta\cos\theta\sin^2\varphi -\\& \frac{M_0}{\gamma}\omega_{c0}\cos\theta\sin\theta - M_0H^{ext}\beta\cos\theta\cos\varphi - M_0 H^{ext} \delta \sin\theta\sin\varphi =0\\
        \frac{\partial U_{mag}}{\partial \varphi} &= \frac{M_0}{\gamma} \omega_{c0}\sin^2\theta\sin\varphi\cos\varphi + M_0\beta\sin\theta\sin\varphi H^{ext} + M_0 H^{ext}\delta \cos\theta\cos\varphi = 0\\
        \frac{\partial U_{mag}}{\partial \delta} &= \frac{M_0}{\gamma} \omega_{ex}\delta + M_0 H^{ext} \cos\theta \sin\varphi = 0
        \end{aligned}
\end{equation}
where the angular frequencies are introduced as follows: $\omega_b = 2K_b\gamma/M_0$, $\omega_{c0} = 2K_c\gamma/M_0$, $\omega_{ex} = 4\lambda_{ex} M_0 \gamma$, $\omega_D = \gamma d / M_0$.\\
Solving the resulting equations we get that equilibrium angles are $\theta_0 = \pi / 2$, $\varphi_0 = 0\text{ or }\pi$, and $\beta_0 = -\omega_D / \omega_{ex} \pm \gamma H^{ext} / \omega_{ex}$, where the $\pm$-sign corresponds to the equilibrium angle $\varphi$: $\varphi_0 = 0$ or $\varphi_0 = \pi$, respectively.

For the small deviations from the equilibrium: $\theta_1= \theta - \theta_0$, $\varphi_1 = \varphi - \varphi_0$, $\delta_1 = \delta - \delta_0$, $\beta_1 = \beta - \beta_0$, $\theta_1\ll1$, $\varphi_1 \ll 1$, $\beta_1 \ll 1$, $\delta_1 \ll 1$, Eq.~(\ref{eq:ml_full}) is linearized as:
\begin{equation}
		\begin{aligned}
			m_x & \approx \mp \beta_0\varphi_1,\,
			m_y  \approx \pm\beta_1,\,
			m_z \approx \delta_1,\\
			l_x& \approx \pm 1,\,
			l_y\approx \pm \varphi_1,\,
			l_z\approx -\theta_1.
		\end{aligned}
		\label{eq:ml_lin}
	\end{equation}
    
In the expression above, $m_y$ is only the dynamic part, without taking $m_{y0} = \pm\beta_0$ into account.

The linearized Lagrangian without constant terms is given by:
\begin{equation}
    \begin{aligned}
    &L = \frac{M_0}{\gamma}(\delta_1\dot{\varphi_1} + \beta_1\dot{\theta_1}) - \frac{M_0}{\gamma}[\omega_{ex}\frac{\delta_1^2 + \beta_1^2}{2} + \omega_{ex}\beta_1\beta_0 - \omega_{ex}\beta_0^2\frac{\theta_1^2}{2} + \omega_b \frac{\theta_1^2}{2} + \omega_{c0} \frac{\theta_1^2 + \varphi_1^2}{2} +\omega_D\beta_1 - \omega_D\beta_0\theta_1^2]\mp\\ &\beta_0\varphi_1 M_0H_x\pm \beta_1 M_0 H_y + \delta_1 M_0 H_z \pm \beta_1 M_0 H^{ext} \mp \beta_0 M_0 \frac{\theta_1^2 + \varphi_1^2}{2}H^{ext} -  A((\nabla\theta_1)^2 + (\nabla \varphi_1)^2) + \frac{1}{8\pi}(\hat{\varepsilon}\textbf{E}^2 - \textbf{B}^2)
    \label{eq:Lagrangian_lin}
\end{aligned}
\end{equation}
Using Euler-Lagrange equation: 
\begin{equation}
    \frac{\partial L}{\partial q_i} - \frac{d}{dt}\left(\frac{\partial L}{\partial \dot{q_i}}\right) - \nabla\cdot \frac{\partial L}{\partial (\nabla q_i)} = 0
    \label{eq:Euler-Lagrange}
\end{equation}
where $q_i$ are the angular variables. Substituting the equilibrium values and retaining linear terms, we obtain:
\begin{equation}
    \begin{aligned}
        \dot{\theta}_1 &= \omega_{ex}\beta_1 \mp \gamma H_y,\\
        \dot{\beta}_1 &= -\omega_a\theta_1
        + \frac{c_m^2}{\omega_{ex}}\nabla^2\theta_1,\\
        \dot{\varphi}_1 &= \omega_{ex}\delta_1 - \gamma H_z,\\
        \dot{\delta}_1 &= -\omega_c\varphi_1
        + \frac{c_m^2}{\omega_{ex}}\nabla^2\varphi_1 \mp \beta_0\gamma H_x.
    \end{aligned}
    \label{eq:full_dynamic}
\end{equation}
Differentiating Eq.~(\ref{eq:full_dynamic}) and introducing damping phenomenologically, we get: 
\begin{equation}
\begin{aligned}
&\ddot{\theta}_1+\kappa_{m2}\dot{\theta}_1+
\omega_{r2}^2\theta_1-c_m^2\partial_x^2\theta_1
=\mp\gamma\dot{H}_y,\\
&\ddot{\beta}_1+\kappa_{m2}\dot{\beta}_1+
\omega_{r2}^2\beta_1-c_m^2\partial_x^2\beta_1
=
\pm\gamma\omega_aH_y \mp\frac{c_m^2}{\omega_{ex}}\gamma\partial_x^2H_y,\\
&\ddot{\varphi}_1+\kappa_{m1}\dot{\varphi}_1+
\omega_{r1}^2\varphi_1-c_m^2\partial_x^2\varphi_1
=-\gamma\dot{H}_z\mp\beta_0\gamma\omega_{ex}H_x,\\
&\ddot{\delta}_1+\kappa_{m1}\dot{\delta}_1+
\omega_{r1}^2\delta_1-c_m^2\partial_x^2\delta_1
=
\gamma\omega_cH_z-\frac{c_m^2}{\omega_{ex}}\gamma\partial_x^2H_z\mp\beta_0\gamma\dot{H_x},\\
&H_x + 4\pi M_0m_x = 0,\\
&\partial_x^2 H_y
-\frac{\varepsilon_z}{c^2}\ddot H_y
=
\frac{4\pi M_0\varepsilon_z}{c^2}\ddot m_y,\\
&\partial_x^2H_z
-\frac{\varepsilon_y}{c^2}\ddot H_z
=
\frac{4\pi M_0\varepsilon_y}{c^2}\ddot m_z,
\end{aligned}
\label{eq:dynamic_diff}
\end{equation}
where $c_m=\sqrt{2\gamma A\omega_{ex}/M_0}$, $\omega_{r1}^2 = \omega_{ex}\omega_c$, and $\omega_{r2}^2 = \omega_{ex}\omega_a$. For the domain considered in the calculations, whose weak-ferromagnetic moment is aligned with the external field, $\omega_c = \omega_{c0} + \gamma H^{ext}/\omega_{ex}\cdot(\omega_D + \gamma H^{ext})$ and $\omega_a = \omega_b + \omega_{c0} + \omega_D/\omega_{ex}\cdot(\omega_D + \gamma H^{ext})$.

Taking into account that solution of the system can be presented as \{$\theta_1, \, \beta_1,\, \varphi_1, \, \delta_1, \, \mathbf{H}$\} $\propto \exp{[i(kx - \omega t)]}$, following equations are obtained:
\begin{equation}
    \begin{aligned}
        \beta_1 &=
\pm
\frac{\gamma H_y
\left(
\omega_a+c_m^2k^2/\omega_{ex}
\right)}
{\omega_{\rm m2}^2(k)-\omega^2-i\kappa_{m2}\omega},
\\[2mm]
\theta_1 &=
\pm
\frac{i\gamma\omega H_y}
{\omega_{\rm m2}^2(k)-\omega^2-i\kappa_{m2}\omega},\\
\delta_1 &= \frac{H_z \gamma (\omega_c + c_m^2k^2/\omega_{ex})\pm i\beta_0\gamma\omega H_x}{\omega_{m1}^2(k) - \omega^2 - i\kappa_{m1}\omega},\\
			\varphi_1 &= \frac{i\omega \gamma H_z \mp \beta_0\gamma\omega_{ex}H_x}{\omega_{m1}^2(k) - \omega^2 - i\kappa_{m1}\omega} .
    \end{aligned}
    \label{eq:dynamic}
\end{equation}
%{\color{red}It should be noted, that from equations (\ref{eq:dynamic}), it is clear that the Dzyaloshinsky constant $d$ determines the relationship between the resonance frequencies $\omega_{r1,2}$ and the external magnetic field $H^{ext}$, and does not affect other dispersion parameters, and consequently, $\mu$. Therefore, it does not directly influence either the magnitude of the splitting or cooperativity. However, if $d$ is small, a large external magnetic field (more than 3 T) will be required to observe anticrossing.}
%\renewcommand{\theequation}{(App.A.\arabic{equation})}

\textbf{B. Yttrium orthoferrite parameters}\\
The calculations used the following yttrium orthoferrite parameters: $M_0 = 596.8\ \text{emu/cm}^3$, exchange constant $\lambda_{ex}= 4200$ ($\omega_{ex} / 2\pi= 28$~THz), inhomogeneous exchange interaction constant $A = 4 \cdot 10^{-7}\ \text{erg/cm}$, gyromagnetic ratio $\gamma = 1.76 \cdot 10^{7}\ \text{1/Oe$\cdot$s}$, Dzyaloshinskii-Moriya interaction constant $d=86.5\cdot10^6 \ \text{erg}~\text{cm}^{-3}$ ($\omega_D/2\pi=0.41$~THz).

Neglecting magneto-optical gyrotropy, the orthorhombic dielectric permittivity tensor has three independent diagonal components. In the coordinates $x\parallel a$, $y\parallel c$, $z\parallel b$ used here,
\[
\hat{\varepsilon}=\mathrm{diag}(\varepsilon_a,\varepsilon_c,\varepsilon_b).
\]
The room-temperature values from Table~3 of \cite{balbashov1995} are $\varepsilon_z=\varepsilon_b=21.9+0.09i$ and $\varepsilon_y=\varepsilon_c=25.4+0.04i$ near the AFM resonance. The component $\varepsilon_x=\varepsilon_a$ is not specified by that table and does not enter the dispersion relations for the two transverse polarizations considered at $\mathbf{k}\parallel a$. No equality between the three diagonal components is imposed by orthorhombic symmetry.

At room temperature: magnetic anisotropy constants $K_b = 87\cdot 10^3~\text{erg}~\text{cm}^{-3}$ and $K_c = 341 \cdot 10^{3}\ \text{erg}~\text{cm}^{-3}$ ($\omega_b / 2\pi = 818$~MHz, $\omega_{c0} / 2\pi = 3.2$~GHz, $\omega_a(0)/ 2\pi = 9.9$~GHz) and the corresponding ferromagnetic (FM) and AFM mode frequencies at zero external magnetic field are $\omega_{r1}/2\pi\ = 299.4   \text{ GHz}$ and $\omega_{r2}/2\pi\ = 526.8   \text{ GHz}$.  At a temperature of 63~K: $K_b = 315\cdot 10^{3}\ \text{erg}~\text{cm}^{-3}$, $K_c = 328\cdot 10^{3}~\text{erg}~\text{cm}^{-3} $ ($\omega_b / 2\pi = 2.95$~GHz, $\omega_{c0} / 2\pi = 3.08$~GHz, $\omega_a(0)/ 2\pi = 11.9$~GHz), and the corresponding FM and AFM mode frequencies at zero external magnetic field are $\omega_{r1}/2\pi\ = 293.7 \text{ GHz}$ and $\omega_{r2}/2\pi\ = 577.9  \text{ GHz}$.

From the dispersion of magneto‑polaritons of the AF and F modes, the diagonal components of the magnetic permeability tensor can be obtained.

\begin{equation}
\begin{aligned}[c]
 \mu_{xx} &= \mu_{aa}=1+\frac{\Delta\mu_a\omega_{r1}^2}{(\omega_{r1}^2-\omega^2-i\omega\kappa_{m1})}, \Delta\mu_a= 4 \pi \chi_{rot}, ~\chi_{rot} = \frac{\omega_M \omega_D^2}{4 \pi \omega_{ex}^2 
\omega_{c0}} \\[4pt]
 \mu_{zz} &= \mu_{bb}=1+\frac{\Delta\mu_b\omega_{r1}^2}{(\omega_{r1}^2-\omega^2-i\omega\kappa_{m1})}, \Delta\mu_b=\frac{\omega_M}{\omega_{ex}} \\[4pt]
 \mu_{yy} &= \mu_{cc}=1+\frac{\Delta\mu_c\omega_{r2}^2}{(\omega_{r2}^2-\omega^2-i\omega\kappa_{m2})}, \Delta\mu_c=\frac{\omega_M}{\omega_{ex}}. \\[4pt]
\end{aligned}
\label{eq:mu}
\end{equation}

\textbf{C. Photonic crystal parameters} 

The AFM resonance frequency shifts from 527 GHz at 300 K to 578 GHz at $\sim 63$ K. For each temperature, the  photonic crystal (PhC) geometry was optimized separately so that the photonic mode remained resonant with the temperature-shifted AFM mode and Fabry-P\'erot modes of the same order were excited: at $T = 300$ $K$ the $\mathrm{YFeO_3}$ plate thickness is $d=470$ $\mu$m, layer thicknesses: $d_{Si} = 39.56$ $\mu$m, $d_{SiO_2} = 75.08$ $\mu$m; at $T = 63$ $K$ the plate thickness is $d=430$ $\mu$m, layer thicknesses: $d_{Si} = 35.76$ $\mu$m, $d_{SiO_2} = 67.86$ $\mu$m. 

\begin{figure*}[h]
    \centering
        \includegraphics[width=0.7\textwidth]{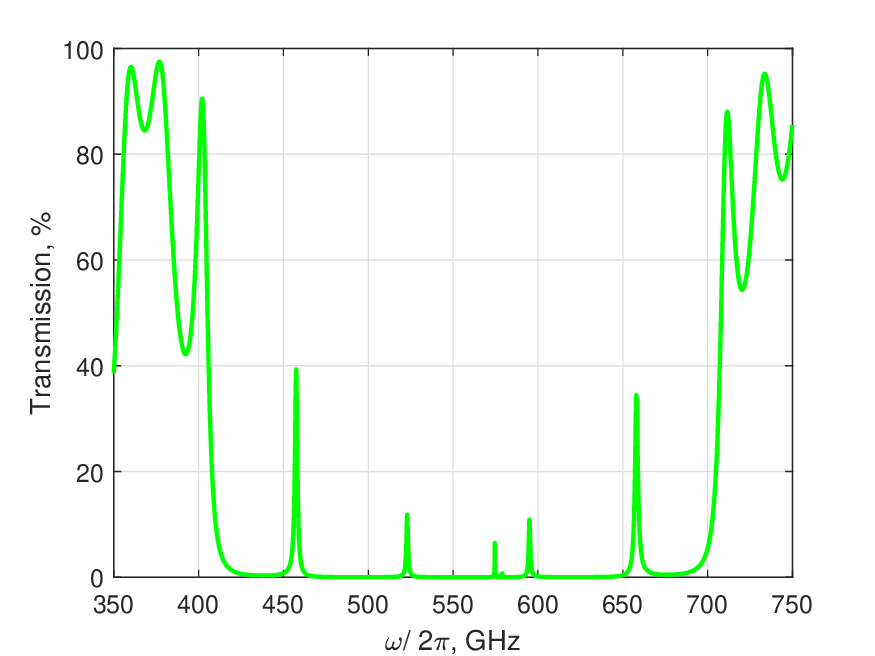}
    \hfill

  \caption{The structure spectrum for a wider frequency range. T = 300 K}
   \label{fig:FM}
\end{figure*}

For the FM mode at 300 K, the PhC was optimized in a similar way at the corresponding FM resonance frequency, close to 299.4 GHz. The parameters of the photonic crystal were: $T = 300$ $K$ the plate thickness is $d=412$ $\mu$m, layer thicknesses: $d_{Si} = 71.28$ $\mu$m, $d_{SiO_2} = 136.11$ $\mu$m.  Material parameters of the PhC layers: $\varepsilon_{Si} = 11.5 + 8\cdot10^{-6}i$, $\varepsilon_{SiO_2} = 3.2 + 2\cdot 10^{-5}i$.

\begin{figure*}[h]
    \centering
        \includegraphics[width=0.7\textwidth]{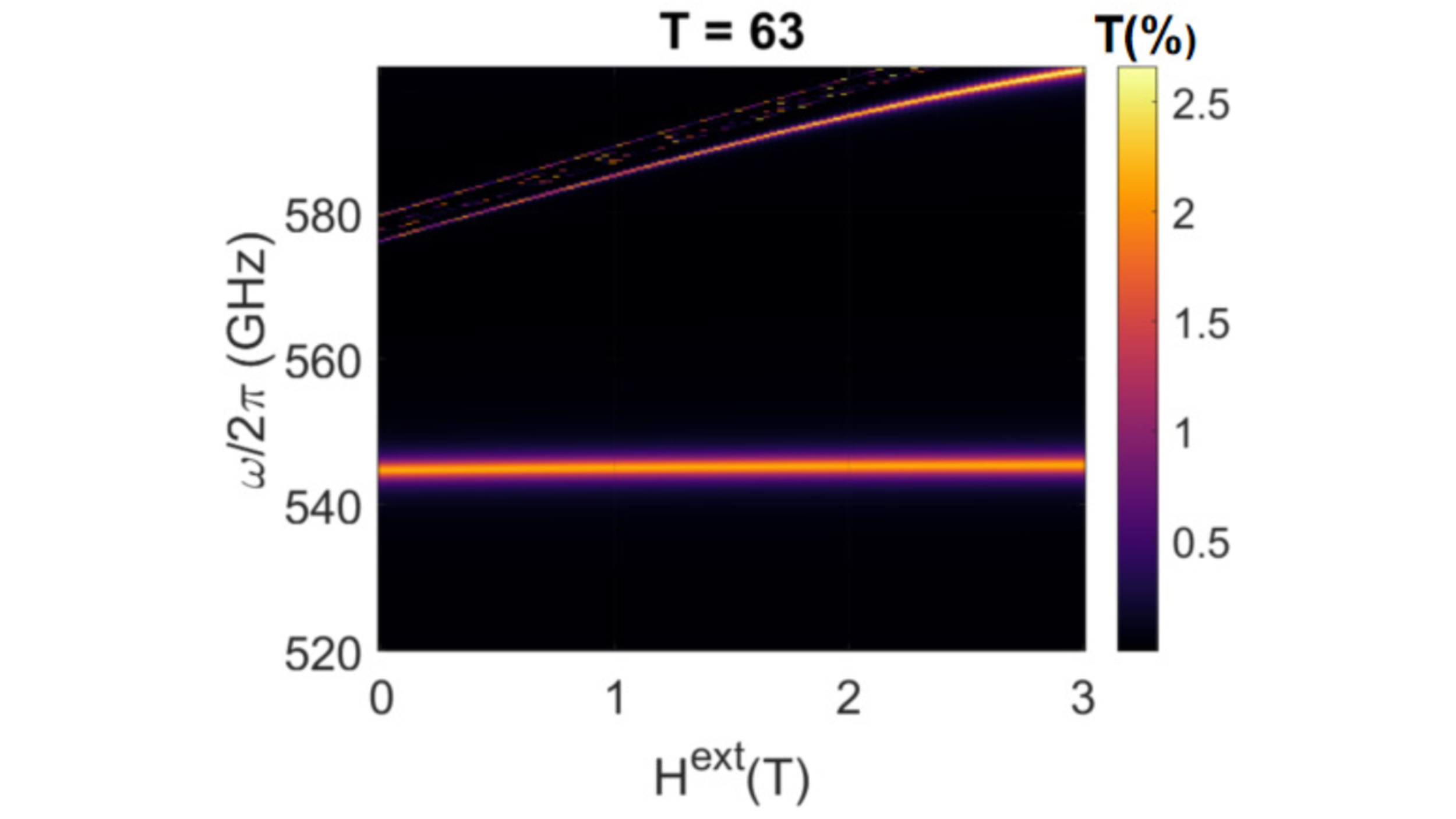}
    \hfill

\caption{Spectrum of a structure with constant geometry with decreasing temperature.}
   \label{fig:DTM}
\end{figure*}

Cooling the sample, optimized for 300 K, down to 63 K shifts the AFM resonance by 33 GHz relative to the photonic one, which leads to the disappearance of a clear anticrossing.(Fig.S\ref{fig:DTM}).

\textbf{D. Transfer matrix method}

Since polarizations do not mix in the configuration considered, each of them can be treated independently with $2\times2$ matrices, and the remaining field components follow from them via Maxwell's equations. The field in a layer is described by a vector of the tangential components $(E_y, H_z)$ (or $(E_z, -H_y)$), which are continuous across interfaces, and is expanded in forward and backward waves. Propagation through a layer is given by a $2\times2$ transfer matrix, and for the whole structure these matrices are multiplied into a total matrix $M$ relating the wave amplitudes in the incident medium and in the substrate. The incident amplitude is given and there is no backward wave in the substrate, which yields a closed system of two equations, from which $r$ and $t$ are expressed through the elements of $M$. Here $r$ and $t$ are the complex ratios of the reflected and transmitted wave amplitudes to the incident one, and $R=|r|^2$ and $T=|t|^2$ are the fractions of the incident power that are reflected and transmitted, 
respectively (for identical media on both sides of the structure).

\textbf{E. Practical requirements for the structure}

Surface roughness and non-parallelism of the plate faces produce a spread $\Delta d$ of the local thickness within the beam spot, which shifts the Fabry--Perot mode by $\delta f = f\,\Delta d/d$ and broadens the photonic branch as $\kappa_{ph}\to\sqrt{\kappa_{ph}^2+(\Delta d\,\omega/d)^2}$. For $f=530$~GHz, $d=440~\mu m$ and $\Delta d<1~\mu m$, $\Delta f<1.2$~GHz, so $\kappa_{ph}$ increases by no more than a factor of $\sqrt{2}$, while $g$ remains practically unchanged (cf. $g$ for the free plate and for the plate in the cavity, Table~I). Scattering losses are negligible. This accuracy is achievable with standard optical polishing.

The large cooperativity is due to the narrow AFM resonance, so the crystal quality is crucial: a polycrystalline sample, whose resonance is many times broader, would substantially reduce it. We therefore consider single-crystal plates cut from a bulk crystal rather than films: at thicknesses of $\sim$0.4~mm they are easier to handle (no substrate removal), whereas thick epitaxial films are difficult to grow.

\textbf{F. Effect of the Dzyaloshinskii--Moriya interaction}

Consider the formal limit $d\to0$ with the anisotropy and exchange constants unchanged. From Eq. \ref{eq:dynamic_diff}, at zero field the mode frequencies are $\omega_{r1}^{2}=\omega_{ex}\omega_{c0}$ and $\omega_{r2}^{2}=\omega_{ex}(\omega_b+\omega_{c0})+\omega_D^{2}$, where $\omega_D=\gamma d/M_0$. Decreasing $d$ therefore lowers only the quasi-AFM mode frequency. For YFeO$_3$ at 300~K it drops from about 527 to 333~GHz, while the quasi-FM mode stays at about 299~GHz. At $d=0$ the equilibrium canting angle is induced solely by the external field, $\beta_0 =\pm \gamma H^{ext} / \omega_{ex}$. The $H_x\beta_0$ term in the equations of motion \ref{eq:dynamic} vanishes at zero field and $d=0$, therefore for propagation along $x$ a wave field initially perpendicular to $x$ remains so. Meanwhile, $g_{\mathrm{AFM}}/2\pi$ decreases from about 7.2 to 4.6~GHz, and $g_{\mathrm{FM}}/2\pi$ remains about 4.1~GHz. Both modes thus still couple to the electromagnetic field without the weak ferromagnetic moment. For unchanged losses and mode overlap, the quasi-AFM cooperativity drops by a factor of about 2.5, and the quasi-FM cooperativity is essentially unaffected.